\documentclass[prl,aps,twocolumn,preprintnumbers]{revtex4}
\usepackage{epsfig,amssymb,amsfonts}

\begin{document}
\title{Quantum Gates Between Flying Qubits via Spin-Independent Scattering}
\date{\today}

\author{Sougato Bose$^1$}
\author{Vladimir Korepin$^2$}
  \address{$^{1}$Department of Physics and Astronomy, University
College London, Gower St., London WC1E 6BT, UK}

    \address{$^{2}$ C. N. Yang Institute for Theoretical Physics, State University of New York at Stony Brook, NY 11794-3840, USA}



\begin{abstract}
We show how the spin independent scattering between two identical flying qubits can be used to implement an entangling quantum gate between them. We consider one dimensional models with a delta interaction in which the qubits undergoing the collision are distinctly labeled by their opposite momenta. The logical states of the qubit may either be two distinct spin (or other internal) states of a fermion or a boson or two distinct momenta magnitudes of a spinless boson. Our scheme could be added to linear optics-like quantum information processing to enhance its efficiency, and can also aid the scaling of quantum computers based on static qubits without resorting to photons. Three distinct ingredients -- the quantum indistinguishability of the qubits, their interaction, and their dimensional confinement, come together in a natural way to enable the quantum gate.
\end{abstract}

\maketitle
\bibliographystyle{apsrev}
Flying qubits are not only required for quantum communication and cryptography, they can also perform quantum computation \cite{KLM,Beenakker,Popescu}. Quantum logic between flying qubits exploits  their indistinguishability and assume them to be {\em mutually non-interacting} -- hence the names ``linear optics" \cite{KLM} and ``free electron" \cite{Beenakker} quantum computation. In fact, for such an approach to be viable one has to engineer circumstances so that the effect of the inter-qubit interactions can be ignored \cite{Popescu}. On the other hand, in the context of photonic qubits, it is known that effective interactions, engineered using atomic or other media, may enhance the efficacy of processing information \cite{Duan-Kimble,Angelakis,Munro,Fleishauer,Gorshkov}. One is thereby motivated to seek similarly efficient quantum information processing (QIP) with material flying qubits which have the advantage of {\em naturally interacting} with each other. Further motivation stems from the fact that for non-interacting mobile fermions, additional ``which-way" detection is necessary for quantum computation \cite{Beenakker} and even for generating entanglement \cite{Bose-Home}, which are not necessarily easy. Thus, if interactions do exist between flying qubits of a given species, one should aim to exploit these for QIP.

 While quantum gates exploiting the mutual interactions of two material flying qubits has not been considered yet, the corresponding situation for {\em static} qubits has been widely studied (e.g., Refs.\cite{loss,jaksch,mandel,deutsch1,deutsch2,philips,grangier}). However, these methods require a precise control of the interaction time of the  qubits or between them and a mediating bus (e.g., Refs.\cite{cirac-zoller,munro-spiller,banchi}). We will instead aim to exploit a much lower control process, namely the scattering of flying qubits. We will {\em fix only the momenta} of the qubits when they are still very far from each other, and then examine whether their scattering provides a useful quantum logic gate. Typically, the quantum gate should only be dictated by the Scattering matrix or {\em S-matrix} acting on the initial state of the two free moving qubits. This is thus an example of {\em minimal control} QIP where nothing other than the initial momenta of the qubits is controlled. Not only will it enable QIP beyond the paradigm of linear optics with material flying qubits, but also potentially connect well separated registers of static qubits. One static qubit from each register should be out-coupled to momenta states in matter wave-guides and made to scatter from each other. The resulting quantum gate will connect separated quantum registers. This may be simpler than interfacing static qubits with photons.

While it is known that both spin-dependent \cite{costa,ciccarello} and spin-independent \cite{lamata,saraga} scattering can entangle, it is highly non-trivial to obtain a useful quantum gate. The amplitudes of reflection and transmission in scattering generally depend on the internal states of the particles involved which makes it difficult to ensure that a unitary operation i.e., a quantum gate acts exclusively on the limited logical (e.g. internal/spin) space that encodes the qubits. Only recently, for non-identical (one static and one mobile) particles, it has been shown that a quantum gate can be engineered from a spin dependent scattering combined with an extra potential \cite{guillermo}. We will show here that one can accomplish a quantum gate merely from the spin independent elastic scattering of two identical particles. This crucially exploits quantum indistinguishablity to
label qubits by their momenta, as well as the equality of the incoming pair and outgoing pair of momenta in one dimension (1D). Our study interfaces QIP and quantum indistinguishability with two other areas, namely the Bethe-Ansatz exact solution of many-body models \cite{Korepin} and the 1D confinement of atoms achieved in recent experiments \cite{Tonks,chips1,chips2,Aspect,Druten-recent}.

\begin{figure}
\centering \includegraphics[width=0.5 \textwidth]{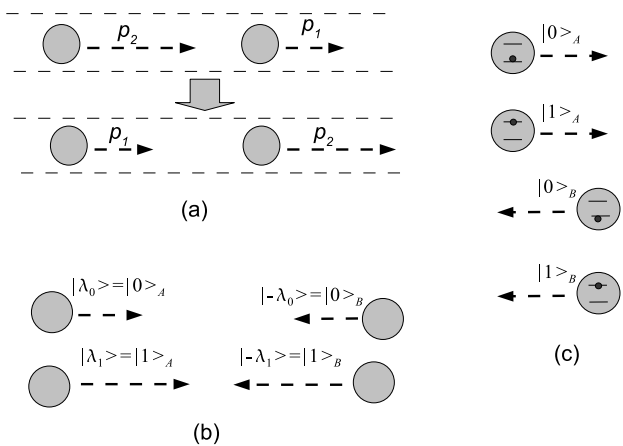}
\caption{Part (a) shows the nature of 1D scattering of two identical particles, whereby incoming particles of momenta $p_1$ and $p_2$ imply outgoing particles of exactly
the same momenta. Part (b) shows the first type of flying qubits we consider, namely encoded in the momenta magnitudes of spinless bosons, where qubits are labeled $A$ or $B$ according to whether they are rightmoving or leftmoving respectively. Part (c) shows the second type (or more conventional type) of flying qubits that we consider, where they are labeled as $A$ and $B$ according to their momenta directions, but it is now their internal states that encode the logical states of the qubit.}
\label{Fig1}
\end{figure}

A two qubit entangling gate is important as it enables universal quantum computation when combined with arbitrary one qubit rotations \cite{bremner}. We consider the spin independent interaction to be a contact interaction between point-like non-relativistic particles. Let us first consider the case of
two {\em spinless} bosons on a line (1D). The Hamiltonian with a delta-function interaction is \cite{Korepin}
\begin{equation}
H=-\frac{\partial^2}{\partial x_1^2}-\frac{\partial^2}{\partial x_2^2}+2 c \delta(x_1-x_2),
\label{ham}
\end{equation}
where $x_1$ and $x_2$ are the coordinates of the two particles. The above model is called the Lieb-Liniger model and has an interesting feature which we will actively exploit. This is the fact that the momenta are individually conserved during scattering. If the incoming particles have momenta $p_1$ and $p_2$, then the outgoing particles {\em also} have momenta $p_1$ and $p_2$ \cite{energy}, as shown in Fig.\ref{Fig1}(a). Thus the scattering matrix is diagonal in the basis of momenta pairs and, is, in fact, only a phase which accumulates on scattering. The eigenfunctions of the model were computed by Lieb and Liniger. The scattering matrix extracted from these wavefunctions is given,
for incident particles with momenta $p_2>p_1$, by \cite{Korepin}
\begin{equation}
S(p_2,p_1)=\frac{p_2-p_1-ic}{p_2-p_1+ic}.
\label{Smat1}
\end{equation}
The phase accumulated on scattering is $-i\ln S(p_2,p_1)$. Note that, as
expected, for non-interacting bosons ($c\rightarrow 0$), their exchange causes no phase change, while when $c\rightarrow \infty$
(impenetrable bosons equivalent to free fermions) and have a $-1$ factor multiplying on exchange.

We now show how the above may be used to accomplish a quantum gate. We consider a frame in which the qubits are moving towards each other so that they eventually collide. Let us call the qubit with momentum towards the {\em right} as qubit $A$, while the qubit with momentum towards the {\em left} is called qubit $B$. As we have chosen the momenta directions to act as the labels of the qubits, and the particles are essentially indistinguishable apart from their momenta, even after the collision, the right-mover is qubit $A$ and the left-mover is qubit $B$. As we are considering spinless bosons, the logical states of the qubits have to be encoded in external degrees of freedom. We consider the logical states to be different {\em magnitudes} of momenta in the {\em same direction}. For qubit $A$, the momenta state $|\lambda_0\rangle$ and $|\lambda_1\rangle$ stand for qubit states $|0\rangle_A$ and $|1\rangle_A$ respectively, while for qubit $B$, the momenta states $|-\lambda_0\rangle$ and $|-\lambda_1\rangle$ stand for qubit states $|0\rangle_B$ and $|1\rangle_B$ respectively (Fig.\ref{Fig1}(b))\cite{second}. We assume $0<\lambda_0<<c$ and $\lambda_1>>c$. Under the above assumptions, the scattering will give rise to the following two qubit quantum gate
\begin{eqnarray}
|0\rangle_A|0\rangle_B \rightarrow (\frac{\lambda_0+\lambda_0-ic}{\lambda_0+\lambda_0+ic}) |0\rangle_A|0\rangle_B=-|0\rangle_A|0\rangle_B \nonumber \\
|0\rangle_A|1\rangle_B \rightarrow (\frac{\lambda_0+\lambda_1-ic}{\lambda_0+\lambda_1+ic}) |0\rangle_A|1\rangle_B=|0\rangle_A|1\rangle_B \nonumber \\
|1\rangle_A|0\rangle_B \rightarrow (\frac{\lambda_1+\lambda_0-ic}{\lambda_1+\lambda_0+ic}) |1\rangle_A|0\rangle_B=|1\rangle_A|0\rangle_B \nonumber \\
|1\rangle_A|1\rangle_B \rightarrow (\frac{\lambda_1+\lambda_1-ic}{\lambda_1+\lambda_1+ic}) |1\rangle_A|1\rangle_B=|1\rangle_A|1\rangle_B.
\label{spinlessgate}
\end{eqnarray}
Essentially, for $|0\rangle_A|0\rangle_B$, the relative momenta of incoming bosons is so small that their interaction strength $c$ is effectively $\infty$ compared to other energies. Thereby they behave as free fermions and obtain a sign change on scattering. For the other three initial qubit states, their relative momenta is so large that they effectively do not see the barriers from each other during collision, and behave as free bosons, thereby giving no sign change.
The above is an example of an entangling two qubit gate (entangles two qubits
initially in the state $\frac{|0\rangle_A+|1\rangle_A}{\sqrt{2}}\frac{|0\rangle_B+|1\rangle_B}{\sqrt{2}}$), and, in fact, is equivalent to a Controlled Z or CZ gate subject to interchanging the $|0\rangle$ and $|1\rangle$ states of the qubit. Note that the form of the qubit that has been presented above has no precedence to our knowledge. The usual form of qubits considered for spinless particles are of time-bin or dual-rail type, which involve distinct paths (i.e., distinct momenta channels
or directions, as in Ref.\cite{Gorshkov}, as opposed to distinct momenta
{\em magnitudes} in the {\em same} channel, as in the case above). As most mechanisms of decoherence act strongly on spatial superpositions, especially for material/charged qubits the momenta magnitude encoding might have some advantage, as long as the momenta difference do not translate to a substantial position difference during the scattering. Of course, for such an encoding, protocols for local unitary operations on a qubit also need to be formulated, for the above entangling gate to be exploitable for universal quantum computation. Without going to details (as we are going to concentrate on other qubit encodings below), we note that multi-photon Bragg scattering of material particles from optical standing waves (wavenumber ${\bf k}$) achieves coherent beamsplitting (without changing any internal states), and have been shown to be able to prepare atoms in momenta superpositions of the form $|p=0\rangle+|p=2n \hbar {\bf k}\rangle$ for $n\sim 12$ \cite{chu} or simply
$(|0\rangle_A+|1\rangle_A)/\sqrt{2}$. 

We next move over to the more traditional form of qubits encoded in an internal degree of freedom of the colliding particles (Fig.\ref{Fig1}(c)). In such cases, the local unitary operations are easy (e.g. laser induced transitions between different atomic internal levels or electronic spin rotations by magnetic fields). The collision is still assumed to have the form of a {\em spin independent} contact (delta) interaction of point-like particles as in Eq.(\ref{ham}). We first consider bosons with two relevant states $|\uparrow\rangle$ and $|\downarrow\rangle$ of some internal degree of freedom (could be any two spin states of a spin-1 boson, for example).
In this case, note that for symmetric states of the internal degrees of freedom, the external degrees of freedom also have to be symmetric and have the {\em same} scattering matrix as spinless bosons (Eq.(\ref{Smat1})). On the other hand, for antisymmetric spin states, the spatial wave function of the two particles is fermionic so that the amplitude for $x_1=x_2$ (the chance of a contact delta interaction) is zero implying that they do not scatter from each other. The above observations lead to the S-matrix (for $p_2>p_1$)
\begin{equation}
S(p_2,p_1)=\frac{(p_2-p_1)-ic\Pi_{12}}{p_2-p_1+ic},
\end{equation}
where $\Pi_{12}$ is the SWAP (permutation) operator acting on the internal (spin) degree of freedom ($\Pi_{12}(|u\rangle_1|v\rangle_2)=|v\rangle_1|u\rangle_2$, where $|u\rangle_1$ and $|v\rangle_2$ are arbitrary spin states of the particles, so that $\Pi_{12}$ is a $4\times 4$ matrix). We define qubits $A$ and $B$ as right moving and left moving as before except that now each qubit is in a definite momenta state, whose magnitudes are $p_A$ and $p_B$ respectively \cite{second}. Thus $p_2=p_A$ and $p_1=-p_B$.
The evolution of the 4 possible qubit states due to the scattering is thereby given by
\begin{eqnarray}
|\uparrow\rangle_A|\uparrow\rangle_B &\rightarrow & (\frac{p_A+p_B-ic}{p_A+p_B+ic}) |\uparrow\rangle_A|\uparrow\rangle_B                                                                                         \nonumber
\\|\downarrow\rangle_A|\downarrow\rangle_B &\rightarrow & (\frac{p_A+p_B-ic}{p_A+p_B+ic}) |\downarrow\rangle_A|\downarrow\rangle_B \nonumber\\
|\uparrow\rangle_A|\downarrow\rangle_B &\rightarrow & \frac{(p_A+p_B)|\uparrow\rangle_A|\downarrow\rangle_B-ic |\downarrow\rangle_A|\uparrow\rangle_B}{p_A+p_B+ic}\nonumber
\\
|\downarrow\rangle_A|\uparrow\rangle_B &\rightarrow &\frac{(p_A+p_B)|\downarrow\rangle_A|\uparrow\rangle_B-ic|\uparrow\rangle_A|\downarrow\rangle_B }{p_A+p_B+ic}
\label{bosongate}
\end{eqnarray}
Unless either $p_A+p_B$ or $c$ vanishes, the above is manifestly an entangling gate, as is evident from the fact that the right hand sides of the last two lines of Eq(\ref{bosongate}) is an entangled state. This gate is most entangling (i.e., the most useful in context of quantum computation, equivalent in usefulness to the well known Controlled NOT or CNOT gate) when $p_A+p_B\approx c$, as then the the right hand sides of the last two lines of Eq(\ref{bosongate}) correspond to maximally entangled states $\frac{e^{-i\frac{\pi}{4}}}{\sqrt{2}}(|\uparrow\rangle_A|\downarrow\rangle_B-i|\downarrow\rangle_A|\uparrow\rangle_B)$ and $\frac{e^{-i\frac{\pi}{4}}}{\sqrt{2}}(|\downarrow\rangle_A|\uparrow\rangle_B-i|\uparrow\rangle_A|\downarrow\rangle_B)$ respectively.

We next consider the case where qubit states are spin states of a spin-1/2 particle (say, electrons or fermionic atoms). This is the conventional encoding in many quantum computation schemes. In this case the $S-$matrix was computed by C. N. Yang \cite{Yang} to be (for $p_2>p_1$)
\begin{equation}
S(p_2,p_1)=\frac{ic\Pi_{12}+p_2-p_1}{ic+p_2-p_1},
\end{equation}
where $\Pi_{12}$ is the SWAP (permutation) operator defined before. As in the previous part of the paper, the qubits themselves are labeled as qubit $A$ and qubit $B$ according to their momenta being right moving of magnitude $p_A$ and left moving of $p_B$ respectively ($p_2=p_A, p_1=-p_B$). Therefore the evolution of spin states is 
\begin{eqnarray}
|\uparrow\rangle_A|\uparrow\rangle_B &\rightarrow&|\uparrow\rangle_A|\uparrow\rangle_B \nonumber
\\|\downarrow\rangle_A|\downarrow\rangle_B &\rightarrow&|\downarrow\rangle_A|\downarrow\rangle_B \nonumber\\
|\uparrow\rangle_A|\downarrow\rangle_B &\rightarrow & \frac{(p_A+p_B)|\uparrow\rangle_A|\downarrow\rangle_B+ic|\downarrow\rangle_A|\uparrow\rangle_B}{ic+p_A+p_B}\nonumber
\\
|\downarrow\rangle_A|\uparrow\rangle_B &\rightarrow & \frac{(p_A+p_B)|\downarrow\rangle_A|\uparrow\rangle_B+ic|\uparrow\rangle_A|\downarrow\rangle_B}{ic+p_A+p_B}
\label{fermiongate}
\end{eqnarray}
Again, as for the bosonic case, for the above gate to produce significant entanglement, $p_A+p_B$ needs to be of the same order as $c$ so that both the terms on the right hand sides of the lower two rows of Eq.(\ref{fermiongate}) are non-negligible. In particular, when $p_A+p_B\approx c$, we have a maximally entangling gate as it converts, e.g.,
$|\uparrow\rangle_A|\downarrow\rangle_B$ to $e^{-i\frac{\pi}{4}}(|\uparrow\rangle_A|\downarrow\rangle_B+i|\downarrow\rangle_A|\uparrow\rangle_B)/\sqrt{2}$. 

 The above gates would aid universal quantum computation by means of scattering with both bosonic and fermionic qubits. The gates of Eqs.(\ref{bosongate}) and (\ref{fermiongate}) are easiest to exploit as the only other requirement, namely local rotations of the qubit states are accomplishable by means of lasers (for atoms) or magnetic fields (for e.g., electrons). Note that the amplitudes in Eqs.(\ref{fermiongate}) and (\ref{bosongate}), e.g.
 $\frac{(p_A+p_B)/c}{i+(p_A+p_B)/c}$,
  are dependent only on the ratio of $(p_A+p_B)/c$ and thereby any spread $\delta p$ of the relative momenta of the incoming particles only affects the amplitudes as $\delta p/c$. Errors can thereby be {\em arbitrarily reduced} in principle by choosing particles with higher $c$. This is opposite to the usual paradigm of gates based on ``timed" interactions, where for a given timing error $\delta t$, stronger interactions enhance the error (while weaker interactions make gates both slower and susceptible to decoherence).

 The most promising implementation of our gates are perhaps with neutral bosonic/fermionic atoms. The delta function interaction we use is, in fact, very realistic and realizable for neutral atoms under strong 1D confinement, as shown by Olshanii \cite{Olshanii}. $^{87}$Rb atoms are suitable, as they have already been strongly confined to 1D atomic waveguides leading to delta interactions \cite{Tonks}. For $^{87}$Rb, with a 3D scattering length $a\sim 50{\AA}$ an axial (for 1D) confinement of $\omega_{\perp}\sim 100$kHz gives (using e.g. Refs.\cite{Olshanii,Busch}) $c\sim 10^6$ m$^{-1}$. Velocities of atoms in 1D waveguides ({\em c.f.} atom lasers \cite{Aspect}) can be mm s$^{-1}$, which translates to $p_A+p_B \sim 10^6$ m$^{-1}$ (here the symbols $p_A$ and $p_B$ have units of wavenumbers). Thereby, $p_A+p_B\approx c$ for {\em optimal gates} is achievable in current technology \cite{comment}. For Eqs.(\ref{bosongate}) and (\ref{fermiongate}), we need state independent waveguides for two spin states -- this requirement has just been met \cite{Druten-recent} in magnetic waveguides (trivially possible in optical waveguides/hollow fibers). Our gate will be an extension of recent collision experiments between different spin species \cite{Zwierlein} with pairs of atoms at a time. Launching exactly two atoms towards each other in 1D should be feasible in atom chips \cite{chips1,chips2} with integrated traps and waveguides and is also a key assumption in many works \cite{Popescu,Calarco}. A proof of principle demonstration of our gates can be made with the technique of Ref.\cite{Calarco} whereby atoms are trapped initially in potential dips inside a larger well and let to roll towards each other in a harmonic potential to acquire their momenta (note that our gate scheme is completely different from Ref.\cite{Calarco}, where the atomic motion is guided by internal states). Lastly, static atomic qubits ({\em c.f.} \cite{grangier}) in well separated traps may be Raman outcoupled \cite{Philips} imparting them momenta towards each other in a 1D waveguide -- this will link distinct registers.

 Ballistic electrons inside carbon nano-tubes or semiconductor nanowires may be another implementation. As electrons are typically screened in a solid, a contact interaction should describe their scattering. Note that the Hubbard model with solely on site interactions models a variety of electronic systems. In a low energy limit, in which the electronic deBroglie wavelength is far larger than the lattice spacing of the Hubbard model, the Hamiltonian reduces to that of Eq.(\ref{ham}).
 
 We have described a mechanism for useful gates between flying qubits using the low control method of scattering. 1D matter waveguides, identical particles and a contact interaction are key requirements for the gate. Exactly solvable models enable the study \cite{Korepin}, while experiments and results on 1D gases \cite{Tonks,chips1,chips2,Aspect,Druten-recent,Olshanii} suggest
 implementations with flying atomic qubits. The scheme may both enhance linear optics-like QIP and connect separated atomic qubit registers.
 
 We thank D. Jaksch, F. Renzoni, D. Schneble and K. Bongs for helpful comments. SB thanks the EPSRC, the Royal Society and the Wolfson Foundation. VK acknowledges financial support by NSF Grant No. DMS-0905744.


\begin{thebibliography}{99}
\bibitem{comment} Even outside this regime, the gates are still entangling and thereby useful unless we are at the free ($p_A+p_B >> c$) or impenetrable ($p_A+p_B<<c$) limits.
\end{thebibliography}
\end{document}